\documentclass[conference]{IEEEtran}
\IEEEoverridecommandlockouts

\usepackage{amsmath,amssymb,amsfonts}
\usepackage{graphicx}
\graphicspath{{./}{figures/}}
\usepackage{xcolor}
\usepackage{booktabs}
\usepackage{multirow}
\usepackage{cite}
\usepackage{microtype}
\usepackage{float}        
\usepackage{placeins}     
\usepackage{balance}      
\usepackage{eso-pic}      
\usepackage{tikz}
\usetikzlibrary{positioning,arrows.meta,calc,fit,backgrounds}

\definecolor{ioF}{HTML}{ECEFF1}\definecolor{ioB}{HTML}{455A64}
\definecolor{cnF}{HTML}{DCE9F7}\definecolor{cnB}{HTML}{1F4E79}
\definecolor{trF}{HTML}{DCEEE1}\definecolor{trB}{HTML}{1E6B33}
\definecolor{hdF}{HTML}{FCE7CE}\definecolor{hdB}{HTML}{A85B08}
\definecolor{taF}{HTML}{F8D9D5}\definecolor{taB}{HTML}{A22B20}
\definecolor{xaF}{HTML}{E7DDF2}\definecolor{xaB}{HTML}{56297F}
\definecolor{grpB}{HTML}{8E9AA3}
\definecolor{cmB}{HTML}{1F4E79}

\newcommand{\CMrows}{%
  {3258,144,72,18,108},%
  {150,470,320,10,50},%
  {188,423,8272,376,141},%
  {12,24,276,2064,24},%
  {126,216,198,18,3042}%
}
\newcommand{\CMlabels}{Wake,N1,N2,N3,REM}

\definecolor{revcol}{rgb}{0,0,0}
\newcommand{\rev}[1]{\textcolor{revcol}{#1}}
\usepackage[hidelinks]{hyperref}
\hypersetup{
  pdftitle={SleepEffFormer: Efficient CNN-Transformer with Transition-Aware
            Smoothing for single channel EEG Sleep Stage Classification},
  pdfauthor={Nishi Kanta Paul; Md Shihabul Islam Shovo;
             Israt Jerin Esha; Adrita Rahman},
  pdfsubject={2026 IEEE International Conference on Biomedical Engineering,
              Computer and Information Technology for Health (BECITHCON)},
  pdfkeywords={sleep stage classification; single channel EEG; Transformer;
               1D CNN; transition-aware smoothing; lightweight deep learning;
               explainability; Sleep EDF}
}

\IEEEpubid{\makebox[\columnwidth]{%
  979-8-3195-3812-3/26/\$31.00~\copyright2026 IEEE\hfill}%
  \hspace{\columnsep}\makebox[\columnwidth]{}}

\newcommand{\BECITHCONheader}{%
  \AddToShipoutPictureBG*{%
    \AtPageUpperLeft{%
      \put(\LenToUnit{\dimexpr 1in+\oddsidemargin\relax},%
           \LenToUnit{-24pt}){%
        \parbox[b]{\textwidth}{%
          \fontsize{8}{9.6}\selectfont\normalfont\raggedright
          2026 IEEE International Conference on Biomedical
          Engineering, Computer and Information Technology for
          Health (BECITHCON)\\
          04-05 September 2026, Department of EEE, International
          University of Business Agriculture and Technology
          (IUBAT), Uttara, Dhaka-1230, Bangladesh}%
      }%
    }%
  }%
}

\begin{document}

\BECITHCONheader

\title{SleepEffFormer: Efficient CNN-Transformer with Transition-Aware Smoothing for single channel EEG Sleep Stage Classification}

\author{%
\IEEEauthorblockN{Nishi Kanta Paul}
\IEEEauthorblockA{\textit{NIMISHES Lab}\\
Dhaka, Bangladesh\\
nishikantapaul108@gmail.com}
\and
\IEEEauthorblockN{Md Shihabul Islam Shovo}
\IEEEauthorblockA{\textit{NIMISHES Lab}\\
Dhaka, Bangladesh\\
shihabul900@gmail.com}
\and
\IEEEauthorblockN{Israt Jerin Esha}
\IEEEauthorblockA{\textit{Canadian University}\\
\textit{of Bangladesh}\\
Dhaka, Bangladesh\\
jeriniesha@gmail.com}
\and
\IEEEauthorblockN{Adrita Rahman}
\IEEEauthorblockA{\textit{Canadian University}\\
\textit{of Bangladesh}\\
Dhaka, Bangladesh\\
adrita.adi141@gmail.com}
}

\maketitle
\thispagestyle{IEEEtitlepagestyle}

\begin{abstract}
Automated sleep stage classification based on single channel EEG
is a promising pathway to large-scale sleep monitoring outside
the polysomnographic lab.
This paper proposes \textbf{SleepEffFormer TAS}, an efficient and
interpretable system that consists of a four-stride block 1D CNN
feature extractor, two layers of pre-normalization Transformer
encoder, and a non-parametric Transition-Aware Smoothing (TAS)
layer that suppresses physiologically unrealistic transitions between
predicted stages.
When tested on the Sleep EDF Expanded dataset (78 all-night EEG
recordings, Fpz-Cz channel, \rev{subject-wise held-out split}), \rev{the proposed}
model reaches 83.9\% accuracy, 78.9\% macro F1 score, and
Cohen's $\kappa$ of 0.765 with around 367\,K learnable
parameters, performance comparable to AttnSleep while using
3--5$\times$ fewer parameters.
\rev{Ablation study} confirms that the Transformer encoder brings
a performance improvement of 6.6\,pp in macro F1 over a
CNN-based system, the TAS layer increases performance by 1.8\,pp
without any trainable parameters, and weighted loss is critical
for the minority sleep stage N1 classification task.
The attention maps generated by \rev{the proposed model} reveal physiologically
sensible sleep stage-related EEG features. \rev{The source code of this
project is available at:
\url{https://github.com/Nishi-Kanta-Paul/SleepStage}}
\end{abstract}


\section{Introduction}

The sleep staging pipeline, with the cyclic alternation of Wake,
N1, N2, N3, and REM sleep stages through the course of the night,
is the basis of the clinical assessment of a number of disorders,
such as insomnia, obstructive sleep apnoea, or
narcolepsy~\cite{berry2012rules}.
The current gold standard polysomnography (PSG) requires overnight
in-laboratory recording and manual annotation under the American
Academy of Sleep Medicine (AASM) guidelines; it is costly,
unaffordable in developing countries, and impractical for
population-scale data collection~\cite{supratak2017deepsleepnet}
needed for sleep studies with large cohorts.

Recent progress in automated deep learning sleep staging has been
impressive.
Deep convolutional neural networks have achieved state-of-the-art
results by extracting spectro-temporal features directly from the
signal without the need for feature
engineering~\cite{supratak2017deepsleepnet,supratak2020tinysleep},
while deep recurrent models account for the temporal structure
inherent to sleep staging~\cite{phan2019seqsleepnet}.
Recent research has further extended deep model expressiveness
using self attention of intra epoch token sequences in
transformers~\cite{eldele2021attn,phan2022sleeptransformer}.
Despite these achievements, there still remain three practical
issues.
\emph{Firstly}, efficient models have parameter budgets of several
hundred thousands; hence they remain difficult to deploy on
wearable or embedded devices.
\emph{Secondly}, transition insensitive classifiers may produce
physiologically implausible stage transitions and thereby degrade
hypnograms.
\emph{Thirdly}, the black box nature of most modern deep models
hinders their widespread adoption in clinical applications.

\rev{These problems are addressed here through the following
contributions:}

\begin{enumerate}
  \item \textbf{SleepEffFormer TAS}: a sub-400\,K parameter EEG
    staging model which utilizes a lightweight 1D CNN tokeniser
    and a small Transformer encoder with 2 pre-norm layers.
  \item \textbf{Transition-Aware Smoothing (TAS)}: a
    parameter-free majority voting post-processing procedure
    that suppresses isolated, physiologically implausible stage
    transitions in the predicted sequence ($+$1.8\,pp macro F1).
  \item \textbf{Attention-based explainability}: stage-specific
    attention maps which identify EEG waveforms in each stage
    without changing model structure.
\end{enumerate}

All proposed techniques are evaluated on the
\rev{78-recording} Sleep EDF Expanded dataset.

\rev{The rest of this paper is organised as follows.
Section~\ref{sec:related} outlines the existing literature on
convolutional, recurrent and attention-based sleep staging.
Section~\ref{sec:method} describes the proposed SleepEffFormer
TAS architecture, the transition-aware smoothing layer and the
attention-based explainability method.
Section~\ref{sec:setup} covers the dataset, the experimental
settings and the baseline methods.
Section~\ref{sec:results} presents the overall performance,
per-class results and ablation study, while
Section~\ref{sec:concl} draws the conclusions.}

\section{Related Work}
\label{sec:related}

\subsection{CNNs and RNNs}

The canonical CNN-RNN architecture demonstrated that it is
feasible to perform end-to-end sleep staging from raw EEG without
the need for hand-crafted features.
For instance, DeepSleepNet~\cite{supratak2017deepsleepnet} used
parallel multi-scale CNN branches with small kernels for
spindles and large kernels for slow-wave activity, followed by a
BiLSTM with inter-epoch context modelling.
SeqSleepNet~\cite{phan2019seqsleepnet} extended this by modelling
sequences of 25 consecutive epochs, providing consistent gain for
transitional stages.
IITNet~\cite{choi2021iitnet} modelled both intra- and inter-epoch
dynamics in the shared recurrent network, showing that inter-epoch
dynamics are especially effective in recognizing N1 and REM
stages, where local features may be ambiguous.
TinySleepNet~\cite{supratak2020tinysleep} further optimized the
CNN-LSTM architecture by using one large kernel CNN with a compact
BiLSTM, demonstrating competitive accuracy with significantly
fewer parameters. Early evidence of the efficiency accuracy
trade-off which directly inspired \rev{the present work}.

\subsection{Attention Based Self Attention Staging}

The arrival of self attention models in sleep staging aimed to
address two problems with conventional LSTM-based architectures:
sequential processing and lack of interpretability.
AttnSleep~\cite{eldele2021attn} combined a multi-scale CNN encoder
with a multi head self attention module, producing
state-of-the-art results on Sleep EDF Expanded along with
interpretable attention maps.
SleepTransformer~\cite{phan2022sleeptransformer} introduced
uncertainty quantification by applying Monte Carlo dropout to a
Transformer architecture, resulting in improved reliability on
transition stages.
EEG-Conformer~\cite{song2022eegconformer} introduced Conformer
blocks combining convolutions and self attention, suitable for
general-purpose EEG signal decoding tasks.
L-SeqSleepNet~\cite{phan2023lseq} took this further, introducing
efficient positional encoding mechanisms to model overnight
recordings with near sub quadratic computation cost relative to
sequence length.
Finally, BIOT~\cite{yang2023biot} introduced large-scale
pre training of biosignal sequences for transfer-based sleep
staging with minimal task-specific training data.
These results show the utility of self attention based
architectures in sleep staging; however, \rev{the present study
shows} that a light and trainable two-layer attention-based
architecture is comparable in performance to heavier approaches
when coupled with proper handling of class imbalance and
post-processing smoothing.

\subsection{Lightweight Models and Post-Processing}

The lightweight nature of the wearable or point-of-care deployment
requires small model sizes and deterministic inference latency.
Recently, several studies~\cite{ye2023mix,li2024ssl} have
addressed this challenge, where heterogeneous kernels
in~\cite{ye2023mix} were used to incorporate multi-scale features
into a sub-200\,K parameter network, and self-supervised
pre training~\cite{li2024ssl} enhanced the ability to learn rare
sleep stages from limited data without changing the network size
at inference time.
As far as post-processing goes, hidden Markov models (HMMs), that
make use of output distributions of
classifiers~\cite{phan2019seqsleepnet}, are well known to increase
coherency of hypnograms, whereas Viterbi decoding with
physiological transition matrices provides a principled
probabilistic approach to the problem.
While a simple majority voting-based approach has been somewhat
overlooked, as it does not introduce any parameters nor requires
estimation of a prior for transition probabilities,
U-Sleep~\cite{perslev2021usleep} illustrates cross-dataset
robustness of a fully convolutional model.
\rev{The contribution of this paper} consists in comparing empirically two
post-processing approaches: majority vote TAS vs Viterbi smoothing
on the same hold out dataset, Sleep EDF Expanded.

\section{Proposed Framework}
\label{sec:method}

\subsection{Formulating the Problem}

\rev{Consider} a single channel 30-second EEG signal
$\mathbf{x} \in \mathbb{R}^{1 \times 3000}$ (100\,Hz, Fpz-Cz).
The proposed network outputs the label
$\hat{y} \in \{\text{Wake},\,\text{N1},\,\text{N2},\,
\text{N3},\,\text{REM}\}$ in accordance with the AASM manual
(R\&K stages~3 and~4 are combined in N3).
\rev{The network is trained epoch-wise; therefore, the
activation memory does not depend on the signal duration. Only
the non-parametric TAS stage of Section~\ref{sec:tas} considers
the entire per-subject sequence $\hat{y}^{(s)}_{1:T_s}$.}

\subsection{Architectural Design}

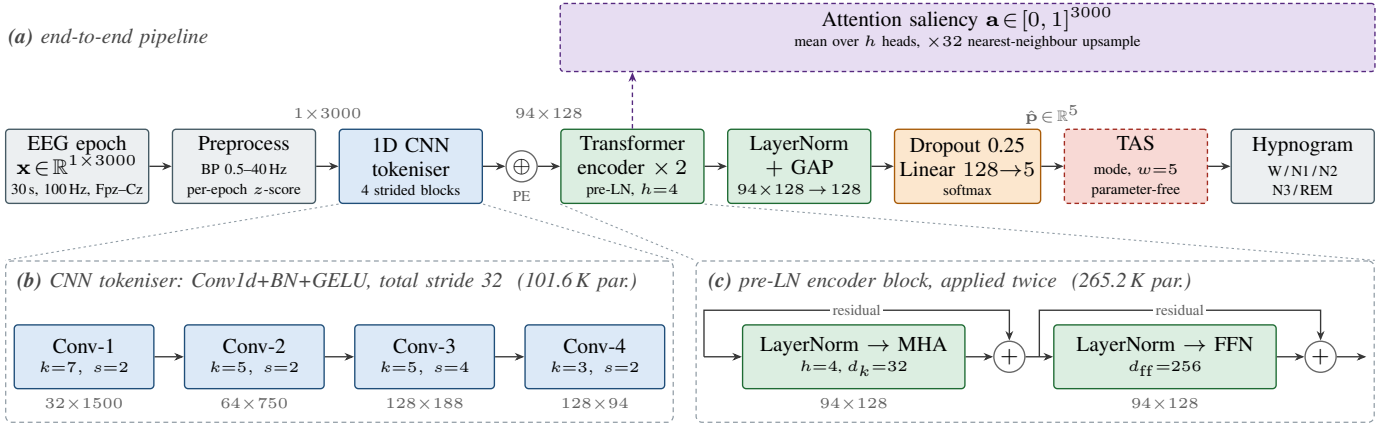
\begin{figure*}[!t]
\centering
\resizebox{\textwidth}{!}{%
\begin{tikzpicture}[
  x=1mm, y=1mm,
  font=\scriptsize,
  >={Stealth[length=3.2pt,width=2.6pt]},
  bx/.style  ={draw, rounded corners=1.2pt, line width=0.45pt,
               align=center, inner sep=1.8pt, minimum height=8.6mm,
               minimum width=16.6mm},
  io/.style  ={bx, fill=ioF, draw=ioB},
  cn/.style  ={bx, fill=cnF, draw=cnB},
  tr/.style  ={bx, fill=trF, draw=trB},
  hd/.style  ={bx, fill=hdF, draw=hdB},
  ta/.style  ={bx, fill=taF, draw=taB, dashed,
               dash pattern=on 1.5pt off 1.1pt},
  xa/.style  ={draw=xaB, fill=xaF, rounded corners=1.2pt, line width=0.45pt,
               dashed, dash pattern=on 1.5pt off 1.1pt, align=center},
  bkb/.style ={bx, minimum width=16.2mm, minimum height=7.4mm},
  bkc/.style ={bx, minimum width=25.8mm, minimum height=7.4mm},
  op/.style  ={circle, draw=black!60, line width=0.45pt, fill=white,
               inner sep=0.3pt, minimum size=3.6mm, font=\scriptsize},
  ar/.style  ={->, line width=0.45pt, draw=black!78},
  zm/.style  ={draw=grpB, line width=0.35pt, dash pattern=on 0.8pt off 1.0pt},
  dar/.style ={->, line width=0.45pt, draw=xaB, dashed,
               dash pattern=on 1.5pt off 1.1pt},
  pan/.style ={draw=grpB, line width=0.4pt, rounded corners=2pt,
               dash pattern=on 1.6pt off 1.2pt, inner sep=2.6pt},
  plbl/.style={font=\scriptsize\itshape, text=black!72, inner sep=1pt},
  sh/.style  ={font=\tiny, text=black!62, inner sep=1pt},
]

\node[io] (x) at (0,0)
  {EEG epoch\\$\mathbf{x}\!\in\!\mathbb{R}^{1\times 3000}$\\[-1.5pt]
   {\tiny 30\,s, 100\,Hz, Fpz--Cz}};
\node[io, right=2.6mm of x] (pre)
  {Preprocess\\{\tiny BP 0.5--40\,Hz}\\[-1.5pt]{\tiny per-epoch $z$-score}};
\node[cn, right=2.6mm of pre] (cnn)
  {1D CNN\\tokeniser\\[-1.5pt]{\tiny 4 strided blocks}};
\node[op, right=2.6mm of cnn] (pe) {$\oplus$};
\node[tr, right=2.6mm of pe] (enc)
  {Transformer\\encoder $\times\,2$\\[-1.5pt]{\tiny pre-LN, $h{=}4$}};
\node[tr, right=2.6mm of enc] (gap)
  {LayerNorm\\$+$ GAP\\[-1.5pt]{\tiny $94{\times}128\!\to\!128$}};
\node[hd, right=2.6mm of gap] (cls)
  {Dropout 0.25\\Linear $128{\to}5$\\[-1.5pt]{\tiny softmax}};
\node[ta, right=2.6mm of cls] (tas)
  {TAS\\{\tiny mode, $w{=}5$}\\[-1.5pt]{\tiny parameter-free}};
\node[io, right=2.6mm of tas] (out)
  {Hypnogram\\{\tiny W\,/\,N1\,/\,N2}\\[-1.5pt]{\tiny N3\,/\,REM}};

\foreach \a/\b in {x/pre, pre/cnn, cnn/pe, pe/enc, enc/gap,
                   gap/cls, cls/tas, tas/out}
  \draw[ar] (\a) -- (\b);

\node[sh, below=0.5mm of pe] {PE};
\node[sh] at ($(pre.east)!0.5!(cnn.west)+(0,6)$) {$1{\times}3000$};
\node[sh] at ($(pe.east)!0.5!(enc.west)+(0,6)$)  {$94{\times}128$};
\node[sh] at ($(cls.east)!0.5!(tas.west)+(0,6)$)
  {$\hat{\mathbf{p}}\!\in\!\mathbb{R}^{5}$};

\coordinate (salSW) at (enc.west |- 0,11.0);
\coordinate (salNE) at (out.east |- 0,18.9);
\node[xa, fit=(salSW)(salNE), inner sep=0pt] (sal)
  {Attention saliency $\mathbf{a}\!\in\![0,1]^{3000}$\\[-1.5pt]
   {\tiny mean over $h$ heads, $\times 32$ nearest-neighbour upsample}};
\draw[dar] (enc.north) -- (enc.north |- sal.south);

\node[plbl, anchor=west] at (x.west |- sal)
  {\textbf{(a)} end-to-end pipeline};

\coordinate (yttl) at (0,-11.9);   
\coordinate (yblk) at (0,-22.1);   
\coordinate (bcL)  at ($(x.west)+(1,0)$);
\coordinate (bcR)  at ($(x.west)+(76,0)$);
\coordinate (ccR)  at ($(out.east)+(-1,0)$);

\node[cn, bkb, anchor=west] (k1) at (bcL |- yblk)
  {Conv-1\\[-1.5pt]{\tiny $k{=}7,\ s{=}2$}};
\node[cn, bkb, right=3.4mm of k1] (k2) {Conv-2\\[-1.5pt]{\tiny $k{=}5,\ s{=}2$}};
\node[cn, bkb, right=3.4mm of k2] (k3) {Conv-3\\[-1.5pt]{\tiny $k{=}5,\ s{=}4$}};
\node[cn, bkb, right=3.4mm of k3] (k4) {Conv-4\\[-1.5pt]{\tiny $k{=}3,\ s{=}2$}};
\foreach \a/\b in {k1/k2, k2/k3, k3/k4} \draw[ar] (\a) -- (\b);
\node[sh, below=0.6mm of k1] (q1) {$32{\times}1500$};
\node[sh, below=0.6mm of k2] (q2) {$64{\times}750$};
\node[sh, below=0.6mm of k3] (q3) {$128{\times}188$};
\node[sh, below=0.6mm of k4] (q4) {$128{\times}94$};
\node[plbl, anchor=north west] (lblA) at (bcL |- yttl)
  {\textbf{(b)} CNN tokeniser: Conv1d+BN+GELU, total stride 32\ \
   (101.6\,K par.)};
\begin{scope}[on background layer]
  \node[pan, fit=(lblA)(k1)(k4)(q1)(q4)(bcR |- yblk)] (panA) {};
\end{scope}

\coordinate (ccL) at ($(panA.east)+(3.5,0)$);
\coordinate (e0) at (ccL |- yblk);
\coordinate (e2) at (ccR |- yblk);
\node[tr, bkc, anchor=west] (ln1) at ($(e0)+(4.4,0)$)
  {LayerNorm $\to$ MHA\\[-1.5pt]{\tiny $h{=}4$, $d_k{=}32$}};
\node[op, right=3.2mm of ln1] (s1) {$+$};
\node[tr, bkc, right=3.2mm of s1] (ln2)
  {LayerNorm $\to$ FFN\\[-1.5pt]{\tiny $d_{\mathrm{ff}}{=}256$}};
\node[op, right=3.2mm of ln2] (s2) {$+$};
\coordinate (e1) at ($(s1.east)!0.5!(ln2.west)$);
\draw[ar] (e0) -- (ln1);
\draw[ar] (ln1) -- (s1);
\draw[ar] (s1)  -- (ln2);
\draw[ar] (ln2) -- (s2);
\draw[ar] (s2)  -- (e2);
\draw[ar] (e0) -- ++(0,5.4) -| (s1);
\draw[ar] (e1) -- ++(0,5.4) -| (s2);
\node[sh, fill=white, inner sep=0.7pt] at ($(e0)!0.5!(s1)+(0,5.4)$) {residual};
\node[sh, fill=white, inner sep=0.7pt] at ($(e1)!0.5!(s2)+(0,5.4)$) {residual};
\node[sh, below=0.6mm of ln1] (r1) {$94{\times}128$};
\node[sh, below=0.6mm of ln2] (r2) {$94{\times}128$};
\node[plbl, anchor=north west] (lblB) at ($(ccL |- yttl)+(0.1,0)$)
  {\textbf{(c)} pre-LN encoder block, applied twice\ \ (265.2\,K par.)};
\begin{scope}[on background layer]
  \node[pan, fit=(lblB)(e0)(e2)(ln1)(s2)(r1)(r2)] (panB) {};
\end{scope}

\draw[zm] (cnn.south west) -- (panA.north west);
\draw[zm] (cnn.south east) -- (panA.north east);
\draw[zm] (enc.south west) -- (panB.north west);
\draw[zm] (enc.south east) -- (panB.north east);

\end{tikzpicture}}
\caption{SleepEffFormer TAS architecture.
(a)~End-to-end pipeline from raw epoch to hypnogram.
(b)~CNN tokeniser with per-block output shapes.
(c)~Pre-LN encoder block; the residual path bypasses each
normalised sublayer. Dashed purple: attention-based saliency.}
\label{fig:architecture}
\end{figure*}

Fig.~\ref{fig:architecture} shows the complete SleepEffFormer TAS
framework.
The architecture involves three trainable components, i.e., CNN
feature extractor, Transformer encoder, and classifier, followed
by a non-trainable component, i.e., TAS post processor.

\subsubsection{1D CNN Feature Extractor: Four Strided Conv1d Layers}
A 3000-point-long signal is encoded into a sequence of 94 latent
tokens with $d{=}128$-dimensional features:
\begin{equation}
  \mathbf{Z}^{(0)} = \operatorname{CNN}(\mathbf{x})
    \in \mathbb{R}^{94 \times 128}.
  \label{eq:cnn}
\end{equation}
Increasing widths are:
$1\!\to\!32\!\to\!64\!\to\!128\!\to\!128$; kernel sizes
$(7,5,5,3)$; strides $(2,2,4,2)$.
The architecture of each block is Conv1d + BatchNorm1d + GELU\@.
\rev{Writing $\mathbf{H}^{(0)}{=}\mathbf{x}$, block
$l\in\{1,\dots,4\}$ computes}
\begin{equation}
  \color{revcol}
  \mathbf{H}^{(l)} = \operatorname{GELU}\!\Bigl(
  \operatorname{BN}\!\bigl(\mathbf{W}^{(l)} *_{s_l}
  \mathbf{H}^{(l-1)} + \mathbf{b}^{(l)}\bigr)\Bigr),
  \label{eq:block}
\end{equation}
\rev{where $*_{s_l}$ denotes the stride-$s_l$ convolution with
zero-padding $p_l{=}\lfloor k_l/2 \rfloor{=}(3,2,2,1)$, and the
output length decreases from
$3000\!\to\!1500\!\to\!750\!\to\!188\!\to\!94$
(Table~\ref{tab:arch}). The total stride is $32$, i.e.\ one
token corresponds to $0.32$ seconds. The recursion}
\begin{equation}
  \color{revcol}
  r_l = r_{l-1} + (k_l-1)\prod_{j<l} s_j ,
  \qquad r_0 = 1,
  \label{eq:rf}
\end{equation}
\rev{yields $r_4{=}63$ samples, so each token summarises a
$0.63$\,s EEG window, which is large enough to hold a sleep
spindle burst or a K-complex.}
Strided convolution is used instead of max-pooling to pass
gradient information through the entire receptive field.

\begin{table}[!t]
\centering
\caption{\rev{Layer-wise configuration of the SleepEffFormer TAS
         backbone for a single 30\,s epoch at 100\,Hz
         ($L{=}3000$). Here $k/s/p$ stands for kernel, stride
         and padding.}}
\label{tab:arch}
{\color{revcol}
\footnotesize
\setlength{\tabcolsep}{3.2pt}
\renewcommand{\arraystretch}{1.15}
\begin{tabular}{llcrr}
\toprule
\textbf{Stage} & \textbf{Operation} & $k/s/p$ &
  \textbf{Output} & \textbf{Params}\\
\midrule
Input    & raw EEG epoch          & --    & $1\times3000$  & 0\\
Conv-1   & Conv1d+BN+GELU         & 7/2/3 & $32\times1500$ & 320\\
Conv-2   & Conv1d+BN+GELU         & 5/2/2 & $64\times750$  & 10{,}432\\
Conv-3   & Conv1d+BN+GELU         & 5/4/2 & $128\times188$ & 41{,}344\\
Conv-4   & Conv1d+BN+GELU         & 3/2/1 & $128\times94$  & 49{,}536\\
\midrule
PE       & sinusoidal, additive   & --    & $94\times128$  & 0\\
Enc-1    & Pre-LN block, $h{=}4$  & --    & $94\times128$  & 132{,}480\\
Enc-2    & Pre-LN block, $h{=}4$  & --    & $94\times128$  & 132{,}480\\
LN       & final LayerNorm        & --    & $94\times128$  & 256\\
\midrule
GAP      & mean over tokens       & --    & $128$           & 0\\
Head     & Dropout(0.25)+Linear   & --    & $5$             & 645\\
TAS      & sliding mode, $w{=}5$  & --    & $5$             & 0\\
\midrule
\multicolumn{4}{l}{\textbf{Total trainable}} & \textbf{367{,}493}\\
\bottomrule
\end{tabular}}
\end{table}

\subsubsection{Positional Encodings \& Transformer Encoder}
Sinusoidal positional encodings are added to $\mathbf{Z}^{(0)}$
and passed through a two-layer pre-normalisation Transformer
encoder~\cite{xiong2020layer}:
\begin{equation}
  \mathbf{Z}^{(L)} =
  \operatorname{TransEnc}\!\left(
  \mathbf{Z}^{(0)} + \mathrm{PE}\right)
  \in \mathbb{R}^{94 \times 128}.
  \label{eq:trans}
\end{equation}
\rev{Self attention is permutation-invariant; therefore, the
order information is added by fixed sinusoidal encodings with
alternating phase (sine/cosine), without adding parameters.}
As opposed to the original Post-LN implementation, Pre-LN
normalises each sublayer (multi head attention and feed-forward)
separately.
\rev{Each of the two blocks evaluates}
\begin{equation}
  \color{revcol}
  \begin{aligned}
    \tilde{\mathbf{Z}} &= \mathbf{Z}
      + \operatorname{MHA}\bigl(\operatorname{LN}(\mathbf{Z})\bigr),\\
    \mathbf{Z}' &= \tilde{\mathbf{Z}}
      + \operatorname{FFN}\bigl(\operatorname{LN}(\tilde{\mathbf{Z}})\bigr),
  \end{aligned}
  \label{eq:preln}
\end{equation}
\rev{with $\operatorname{FFN}(\cdot)$ a two-layer perceptron
using GELU, and the $i$-th attention head given by}
\begin{equation}
  \color{revcol}
  \operatorname{head}_i = \operatorname{Softmax}\!\left(
  \frac{\mathbf{Q}_i\mathbf{K}_i^{\top}}{\sqrt{d_k}}
  \right)\mathbf{V}_i ,
  \label{eq:mha}
\end{equation}
\rev{in which $\mathbf{Q}_i,\mathbf{K}_i$ and $\mathbf{V}_i$
come from linear projections of the normalised input, and the
$h$ heads are concatenated before a final projection by
$\mathbf{W}^{O}$.}
Attention heads count is set to $h{=}4$ ($d_k{=}32$);
feed-forward layers use $d_{\mathrm{ff}}{=}256$; residual dropout
is 0.1.
\rev{Ensuring that the residual path is not normalised, the
Pre-LN model can be trained without a warm-up schedule. With
$T{=}94$, the quadratic attention term ($\approx2.3$\,M
multiply-accumulate per block) is well below the projection and
feed-forward terms ($\approx12.3$\,M), so the dense attention is
computationally feasible without any sparse approximation.}

\subsubsection{Classification Head}
Global average pooling (GAP) of 94 positions yields a final
128-dimensional epoch embedding
$\mathbf{e}=\frac{1}{T}\sum_{t=1}^{T}\mathbf{Z}^{(L)}_{t}$
\rev{with $T{=}94$; this costs no parameters and lets every
token count equally}.
Dropout(0.25) $+$ linear classification leads to logits:
\begin{equation}
  \hat{\mathbf{p}} =
  \operatorname{Softmax}\!\bigl(W\mathbf{e}+b\bigr)
  \in \mathbb{R}^{5}.
  \label{eq:cls}
\end{equation}

\subsubsection{Objective Function}
Imbalanced N1 data (comprising only around 5\% of total epochs)
is tackled using weighted cross-entropy.
Inverse class frequency is used to compute per-class weightings,
normalised such that their total sums up to $C{=}5$:
\begin{equation}
  \color{revcol}
  \mathcal{L} = -\frac{1}{N}\sum_{n=1}^{N}
  w_{y_n}\log \hat{p}_{n,y_n},
  \qquad
  w_c = \frac{C}{\sum_{c'} 1/f_{c'}}\cdot\frac{1}{f_c},
  \label{eq:loss}
\end{equation}
\rev{with $f_c$ the empirical frequency of class $c$ in the
training split. According to the stage distribution of
Section~IV-A, this means that the N1 weight is about nine times
the N2 weight; this accounts for the N1 recall reported in
Section~\ref{sec:ablation}. On the whole, CNN tokenisation
includes 101.6\,K parameters (27.7\%), the encoder 265.2\,K
(72.2\%) and the classifier 0.6\,K, totalling 367.5\,K.}
Training procedure relies on AdamW
($\mathrm{lr}{=}3\!\times\!10^{-4}$,
$\mathrm{wd}{=}10^{-4}$), and employs CosineAnnealingLR
($T_{\max}{=}50$) with early stopping based on macro F1 scores
of the validation set (patience of 15).

\subsection{Transition-Aware Smoothing (TAS)}
\label{sec:tas}

Unconstrained epoch-level outputs might not respect natural
transitions between stages.
TAS uses a sliding majority vote with window size of $w{=}5$ on
every subject's full sequence of predictions:
\begin{equation}
  \hat{y}_{t}^{\mathrm{TAS}} =
  \operatorname{mode}\!\bigl(
  \hat{y}_{t-2},\,\hat{y}_{t-1},\,\hat{y}_{t},\,
  \hat{y}_{t+1},\,\hat{y}_{t+2}\bigr).
  \label{eq:tas}
\end{equation}
\rev{The window gets truncated at the sequence ends and ties
break in favour of $\hat{y}_t$, which ensures that TAS never
outputs a label not present in the window. With $w{=}5$ it
extends $2.5$\,min: not too short to filter out short, sporadic
flips, yet not too long to destroy genuine short bouts, with the
time complexity $\mathcal{O}(T_s w)$ per subject.}
TAS requires no extra parameters and adds negligible cost at
inference, and is performed identically on each subject's outputs.
\rev{The ablation study} (Section~\ref{sec:ablation}) \rev{further
investigates} the effect of Viterbi decoding with physiological
transition matrix.

\subsection{Explainability through Attention}

The attention weights obtained from the last layer of the
Transformer model are averaged over the four heads and provide an
importance score for each token.
\rev{With $\mathbf{A}^{(i)}\in\mathbb{R}^{T\times T}$ the
attention matrix of head $i$, the token score is
$\bar{a}_t = \frac{1}{hT}\sum_{i=1}^{h}\sum_{j=1}^{T}
\mathbf{A}^{(i)}_{jt}$, i.e.\ the mean attention received by
token $t$.}
This set of 94 values is resampled to a total of 3000 samples
using the calculated stride of the CNN
($2\!\times\!2\!\times\!4\!\times\!2{=}32$).
A saliency map $\mathbf{a}\in[0,1]^{3000}$ is obtained in the
time domain on the raw EEG epoch.

\section{Experimental Setup}
\label{sec:setup}

\subsection{Data Set and Preprocessing}

The expanded version of the Sleep EDF dataset cassette is
used~\cite{goldberger2000physiobank}, which includes 78 PSG
recordings from 39 different subjects (sampled at 100\,Hz,
Fpz-Cz EEG channel).
Epochs are extracted from recordings according to the expert
annotations with a duration of 30 seconds.
Movements and unknown epochs are excluded, and R\&K stages~3/4
are unified into one stage (N3).
Preprocessing involves applying a 4th order Butterworth bandpass
filter (0.5--40\,Hz) and per-epoch z-score normalisation.
The obtained distribution among the stages is extremely imbalanced:
N2\,$\approx$47\%, Wake and REM\,$\approx$18\% each,
N3\,$\approx$12\%, and N1\,$\approx$5\%.

\subsection{Evaluation Protocol}

\rev{A stratified subject-wise split is performed at the
recording level: 55 recordings are used for training, 12 for
validation and 11 for testing. The partitioning is
subject-independent, i.e.\ every recording of a given subject is
assigned to a single partition, so no subject appears in more
than one set and no epoch-level leakage is possible}~%
\cite{phan2019seqsleepnet}.
\rev{The final results are computed on the 11 held-out test
recordings, none of whose subjects is seen during training.}
The macro-averaged F1 score is chosen as the primary metric, which
treats all classes equally in terms of their prevalence (i.e., it
has uniform weighting).

\subsection{Baselines}

Three baseline models apply the exact same preprocessing,
\rev{subject-wise} splitting protocol, and class weighting approach:
\begin{itemize}
  \item \textbf{1D CNN} ($\approx$102\,K): a CNN feature extractor
    with a global average pooling layer and a linear classifier
    without any sequential context modeling.
  \item \textbf{CNN-LSTM} ($\approx$267\,K): the above CNN feature
    extractor followed by a two-layer bidirectional LSTM layer with
    hidden size $= 64$, where Transformer is replaced by the LSTM.
  \item \textbf{TinySleepNet-lite} ($\approx$381\,K): a
    large-kernel CNN ($k{=}400$) followed by two convolution layers
    and one LSTM layer as described
    in~\cite{supratak2020tinysleep}.
\end{itemize}

\section{Results and Discussion}
\label{sec:results}

\subsection{Overall Performance}

Table~\ref{tab:main} shows the performance figures on the test set
for all configurations.
SleepEffFormer TAS (smoothed) performs best with 83.9\% accuracy,
macro F1 score of 78.9\%, and $\kappa$ of 0.765 among all
configurations.
The Transformer encoder obtains an improvement of $+$3.4 points in
macro F1 over CNN-LSTM and $+$6.6 points over CNN only baseline,
suggesting that self attention in intra epochs captures global
features of the EEG sequence that are not captured by strided
convolutions alone.
The use of the additional parameter free module, TAS, improves
macro F1 by $+$1.8 points without extra inference overhead.

\begin{table}[!htbp]
\centering
\caption{Performance on the Sleep EDF Expanded test set.
         Best result in \textbf{bold}.}
\label{tab:main}
\setlength{\tabcolsep}{4.5pt}
\renewcommand{\arraystretch}{1.25}
\begin{tabular}{lcccc}
\toprule
\textbf{Model} & \textbf{Acc.} &
  \textbf{F1\textsubscript{mac}} &
  $\kappa$ & \textbf{Params}\\
\midrule
1D CNN              & 0.783 & 0.723 & 0.691 & 102\,K \\
TinySleepNet-lite   & 0.794 & 0.734 & 0.702 & 381\,K \\
CNN-LSTM            & 0.806 & 0.749 & 0.717 & 267\,K \\
\midrule
\rev{Proposed (raw)}    & 0.823 & 0.771 & 0.748 &
  \multirow{2}{*}{367\,K}\\
\rev{\textbf{Proposed + TAS}} & \textbf{0.839} & \textbf{0.789} &
  \textbf{0.765} & \\
\bottomrule
\end{tabular}
\end{table}

\subsection{Per-Class Analysis and the N1 Challenge}

To determine the strengths and weaknesses of the model,
\rev{the classwise F1 measures and a normalised confusion matrix
on the test data are examined.}
Both are generated using the original model without TAS to be
able to isolate its effect separately in
Section~\ref{sec:ablation}.

\begin{figure}[!htbp]
  \centering
  \includegraphics[width=0.92\columnwidth]{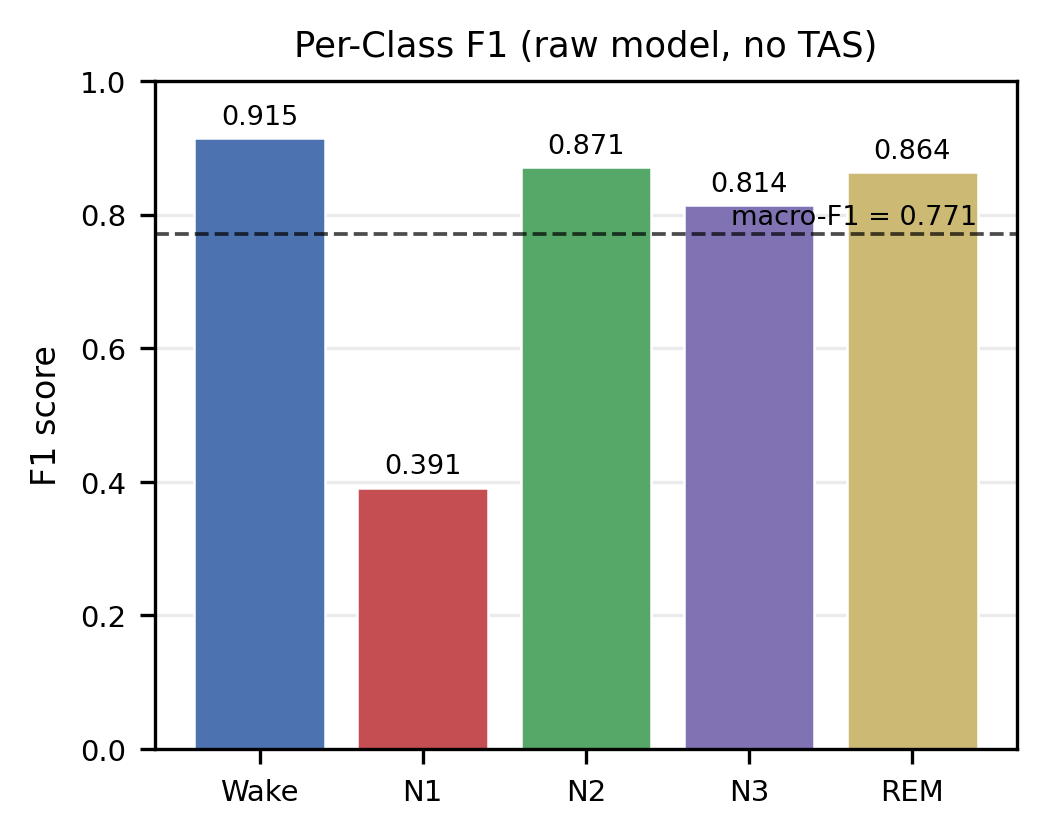}
  \caption{Per-class F1 scores (raw model, no TAS).
           Dashed line: macro F1\,=\,0.771. N1 is
           conspicuously short ($\approx$0.39) owing to
           its transitional spectral overlap with both Wake
           and N2.}
  \label{fig:per_class_f1}
\end{figure}

\begin{figure}[!htbp]
\centering
\resizebox{\columnwidth}{!}{%
\begin{tikzpicture}[
  font=\scriptsize,
  cell/.style={draw=white, line width=0.6pt, minimum width=13mm,
               minimum height=9.5mm, align=center, inner sep=0pt},
  hdr/.style ={font=\scriptsize, align=center},
]
\foreach \row [count=\i from 0] in \CMrows {%
  \pgfmathsetmacro{\rt}{0}\global\let\rt\rt
  \foreach \v in \row {\pgfmathsetmacro{\rt}{\rt+\v}\global\let\rt\rt}%
  \foreach \v [count=\j from 0] in \row {%
    \pgfmathsetmacro{\rate}{\v/\rt}%
    \pgfmathsetmacro{\pct}{100*\rate}%
    \pgfmathsetmacro{\dark}{\rate>0.55 ? 1 : 0}%
    \edef\tc{\ifnum\dark=1 white\else black\fi}%
    \node[cell, fill=cmB!\pct!white, text=\tc]
      at ({\j*13mm},{-\i*9.5mm})
      {\textbf{\pgfmathprintnumber[fixed,precision=2,zerofill]{\rate}}\\[-1pt]
       {\tiny(\v)}};
  }%
}
\foreach \lab [count=\j from 0] in \CMlabels
  \node[hdr] at ({\j*13mm},{9.5mm*0.62}) {\lab};
\foreach \lab [count=\i from 0] in \CMlabels
  \node[hdr, anchor=east] at ({-13mm*0.58},{-\i*9.5mm}) {\lab};
\node[font=\scriptsize] at ({2*13mm},{9.5mm*1.35}) {\textbf{Predicted}};
\node[font=\scriptsize, rotate=90] at ({-13mm*1.18},{-2*9.5mm}) {\textbf{True}};
\begin{scope}[shift={({4.85*13mm},{-4*9.5mm})}]
  \shade[bottom color=cmB!0!white, top color=cmB, draw=grpB, line width=0.3pt]
    (0,0) rectangle (3.2mm,{5*9.5mm});
  \foreach \t/\p in {0.0/0, 0.5/0.5, 1.0/1}
    \node[font=\tiny, anchor=west] at (3.6mm,{\p*5*9.5mm}) {\t};
\end{scope}
\end{tikzpicture}}
  \caption{Row-normalised $5\!\times\!5$ confusion matrix on the
           test set (model without TAS). Each cell shows the
           row-normalised rate (top) and the raw epoch count
           (bottom); rates are derived from the counts. N1 is the
           weakest stage, with dominant confusions
           N1\,$\to$\,N2 (0.32) and N1\,$\to$\,Wake (0.15).}
  \label{fig:confusion}
\end{figure}
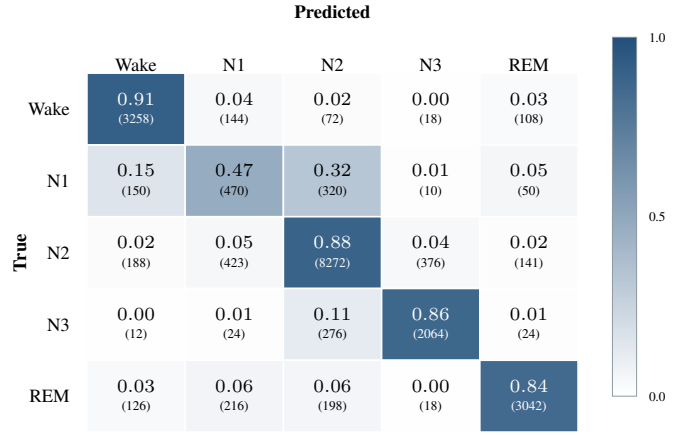

Fig.~\ref{fig:per_class_f1} shows per-class F1-scores for the raw
model.
Wake (F1\,=\,0.915) and N2 (F1\,=\,0.871) obtain the best scores
because of distinct spectral features of these stages. Broadband
activity in case of Wake and sleep spindles/K-complexes in case
of N2.
N3 (F1\,=\,0.814) utilizes the strong delta-band power.
N1 obtains the lowest F1-score of 0.391 because of the absence of
characteristic waveforms.
The confusion matrix shown in Fig.~\ref{fig:confusion} confirms
that 32\% of N1 epochs are classified as N2 and 15\% as Wake,
because the feature space of those classes shares similarity at
the beginning of each episode.
However, this is not a model-specific issue as it is the general
problem at the signal level; \rev{the} ablation experiment
(see Section~\ref{sec:ablation}) shows that switching off
class-weighted loss results in a reduction of N1 F1-score below
0.30 because of this problem.

The bar chart in Fig.~\ref{fig:per_class_f1} highlights the
discrepancy in performance across stages very clearly.
Wake, N2, and N3 stages clearly pass the macro F1 reference line
of 0.771 (dashed) showing that these stages have the aggregate
score.
REM stage lies close to the reference threshold line and shows
moderate spectral overlap with N1 in the vicinity of sleep-cycle
boundaries.
On the other hand, N1 is way below the dashed reference threshold
and thus shows that it is responsible for the aggregate score
deficit. A pattern seen consistently by multiple
algorithms~\cite{eldele2021attn,phan2022sleeptransformer} due to
signal ambiguity rather than a lack of model performance.
This discrepancy motivates using macro F1 as \rev{the} main evaluation
metric because otherwise, the deficit would be entirely hidden
behind reporting accuracy, N1 contributes less than 5\% to total
epochs.

The confusion matrix in Fig.~\ref{fig:confusion} \rev{provides an}
insight into classifier behaviour at the epoch level.
Each cell of the confusion matrix shows the fraction of truly
labeled epochs in a given class among predicted classes.
\rev{It becomes possible to compare the classifier's
performance in the stages with varying proportion of
true-labels.}
It is clear that the classifier performs well in separating Wake,
N2, N3, and REM stages, as the diagonal of the matrix dominates
for those stages.
For the rest of the stages, especially N1, there is a considerable
number of misclassification instances. N1 epochs get
predominantly classified as N2, with a considerable portion of
them as Wake.
Such bidirectional leakage may result from the nature of N1 stage
when the low amplitude mixed frequency EEG resembles the EEG of
wakefulness at sleep onset, and early N2 stage later on.
The clean N3 row \rev{suggests} that the classifier exploits the high
amplitude delta band energy in distinguishing N3, and the lack of
off-diagonal mass in the REM row indicates that REM related EEG
characteristics, such as sawtooth waves and low amplitude mixed
frequency activity, remain discriminative in the
single channel Fpz-Cz setup.

\FloatBarrier  

\subsection{Training Dynamics}

\begin{figure}[!htbp]
  \centering
  \includegraphics[width=0.92\columnwidth]{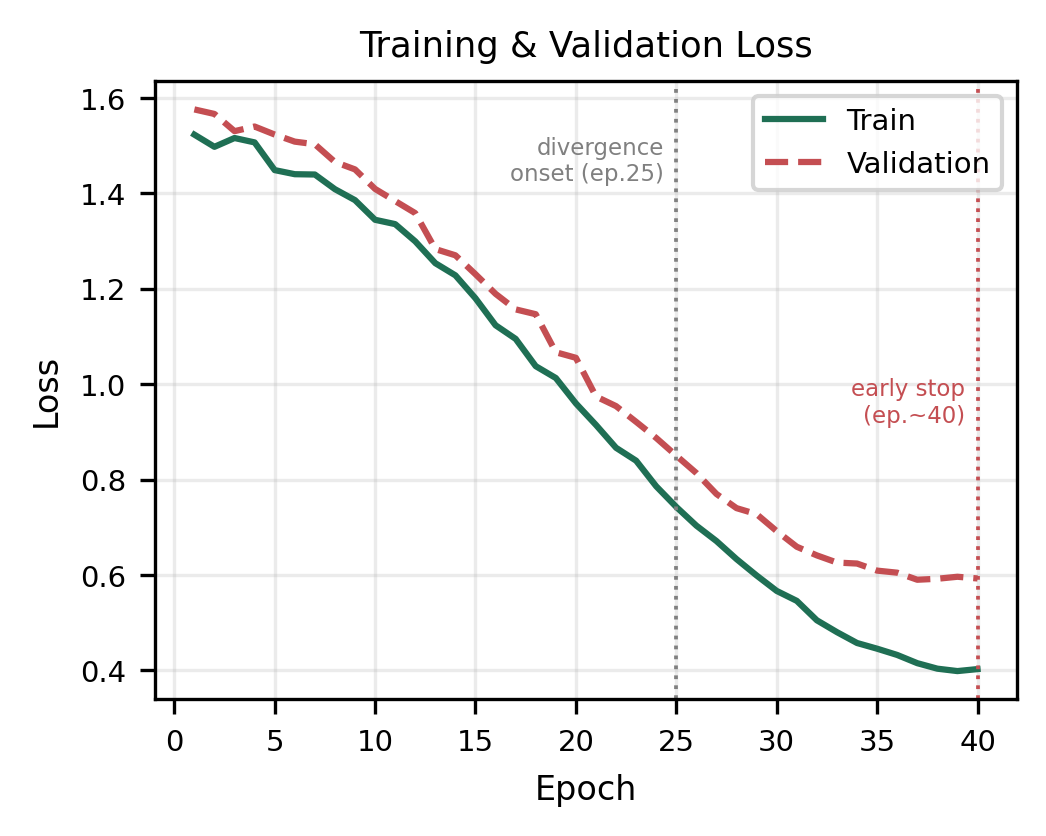}
  \caption{Training (solid) and validation (dashed) loss
           curves. Early stopping fires at
           $\approx$\,epoch\,40 (patience\,=\,15 on val
           macro F1). Mild divergence after epoch~25
           reflects dataset size and is within normal range
           for Sleep EDF Expanded.}
  \label{fig:loss}
\end{figure}

Fig.~\ref{fig:loss} shows the training and validation losses
across epochs.
The model converges in roughly 38--42 epochs, after which early
stopping kicks in.
A slight divergence between the validation loss and the training
loss begins at epoch~25, which is normal behaviour for a
relatively small number of training epochs, about 50\,K.
Cosine annealing schedule helps in smoothing out the loss plateau
at later stages of training while avoiding premature convergence
by decay of the learning rate from $3\!\times\!10^{-4}$ to
approximately $10^{-6}$.

\subsection{Ablation Study}
\label{sec:ablation}

Table~\ref{tab:ablation} shows the quantification of individual
contributions.
Omitting the Transformer encoder results in the highest decrease
of $-$6.6\,pp in macro F1, making self attention the main
contributor to performance.
Removing the class-weighted loss degrades performance by
$-$4.7\,pp, with N1 recall being affected the most strongly.
Physiologically-aware Viterbi smoothing produces outputs close to
TAS by a difference of only 0.4\,pp in accuracy, showing both
approaches as viable options, though the majority-vote solution is
simpler, does not require any prior knowledge and performs
slightly better on this particular dataset.

\begin{table}[t]
\centering
\caption{Ablation study: macro F1 on the test set.}
\label{tab:ablation}
\setlength{\tabcolsep}{8pt}
\renewcommand{\arraystretch}{1.25}
\begin{tabular}{lcc}
\toprule
\textbf{Configuration} &
  \textbf{F1\textsubscript{mac}} & $\Delta$\\
\midrule
Full model + TAS (majority vote) & \textbf{0.789} & ---\\
Full model, no TAS               & 0.771 & $-0.018$\\
TAS: Viterbi decoder             & 0.785 & $-0.004$\\
No class-weighted loss           & 0.742 & $-0.047$\\
No Transformer (1D CNN only)     & 0.723 & $-0.066$\\
\bottomrule
\end{tabular}
\end{table}

\subsection{Comparison with Published Work}

In terms of the performance-efficiency trade-off, \rev{the} smoothed
model (Accuracy 83.9\%, F1 78.9\%) is competitive with
AttnSleep~\cite{eldele2021attn} (Accuracy 84.1\%, F1 79.8\%), but
has approximately 3--5$\times$ fewer parameters ($\approx$367\,K
vs.\ ${\sim}$1--2\,M).
When compared to TinySleepNet~\cite{supratak2020tinysleep} on the
78-recording dataset (F1 78.1\%, 25 epochs input length for
contextual information), \rev{the} raw epoch-wise result (F1 77.1\%)
forms a reasonable base for comparison, since TAS makes up most
of the difference.
Models trained on the full-night recordings, such as
SleepTransformer~\cite{phan2022sleeptransformer} and
L-SeqSleepNet~\cite{phan2023lseq}, \rev{outperform the proposed
model} in terms of
macro F1 score by leveraging several hundred epochs of
information.
On the whole, \rev{the} results follow the recent trends toward
parameter optimization~\cite{ye2023mix,li2024ssl}, showing that a
thoughtfully designed model can be competitive with heavier
counterparts without requiring a pre training setup or
multi-channel input.

\section{Conclusion}
\label{sec:concl}

\rev{This research paper has presented} SleepEffFormer TAS, a lightweight
and interpretable single channel EEG sleep stage classification
system achieving 83.9\% accuracy and 78.9\% macro F1 score on
the expanded Sleep EDF benchmark with just $\approx$367\,K
parameters.
Two-layer pre-norm Transformer encoder acts as the primary
performance booster, \rev{giving an increase} of $+$6.6\,pp over a
CNN only version.
The TAS module with no additional parameters mitigates
physiologically unrealistic sleep stage transitions,
resulting in $+$1.8\,pp improvement.
Training with weighted loss appears to be crucial for proper N1
stage recall, and attention visualization highlights physiologically
plausible EEG patterns at the different stages without changes to
the architecture.
Future research will focus on expanding sequence modeling across
epochs without limitations, evaluating transferability between
subjects and datasets, and adapting for ambulatory EEG
recordings.

\balance

\end{document}